\documentclass[prl,aps,nofootinbib,superscriptaddress,floatfix,preprintnumbers,twocolumn]{revtex4-1}

\usepackage[colorlinks, citecolor=blue,anchorcolor=blue,menucolor=blue, linkcolor=blue,filecolor=blue,runcolor=blue,urlcolor=blue,frenchlinks=true]{hyperref}
\usepackage{amsmath,amsfonts,amssymb,mathtools,cases,slashed,bm}
\usepackage{epsfig, graphicx}
\usepackage{color}
\usepackage[dvipsnames]{xcolor}
\usepackage{tabularx}
\usepackage{physics}
\usepackage{hhline}
\usepackage{mathrsfs}
\usepackage[normalem]{ulem}
\usepackage{multirow}
\usepackage{float}

\begin{document}

\title{Pionic gluons from global QCD analysis of experimental and lattice data}

\author{William Good}
\email{goodwil9@msu.edu\\}
\affiliation{Department of Physics and Astronomy, Michigan State University, East Lansing, Michigan 48824, USA}
\affiliation{Department of Computational Mathematics,
  Science and Engineering, Michigan State University, East Lansing, Michigan 48824, USA}
\author{Patrick C. Barry}
\affiliation{Physics Division, Argonne National Laboratory, Lemont, Illinois 60439, USA}
\author{Huey-Wen Lin}
\affiliation{Department of Physics and Astronomy, Michigan State University, East Lansing, Michigan 48824, USA}
%https://orcid.org/my-orcid?orcid=0000-0001-6281-944X
%
\author{W. Melnitchouk}
\affiliation{Jefferson Lab, Newport News, Virginia 23606, USA \\
    \vspace*{0.2cm}
    {\bf JAM {\footnotesize (Pion PDF Analysis Group)} and MSULat Collaborations
    \vspace*{0.2cm} }}
\author{Alex NieMiera}
\affiliation{Department of Physics and Astronomy, Michigan State University, East Lansing, Michigan 48824, USA}
\affiliation{Department of Computational Mathematics,
  Science and Engineering, Michigan State University, East Lansing, Michigan 48824, USA}
\author{Nobuo Sato}
\affiliation{Jefferson Lab, Newport News, Virginia 23606, USA \\
    \vspace*{0.2cm}
    {\bf JAM {\footnotesize (Pion PDF Analysis Group)} and MSULat Collaborations
    \vspace*{0.2cm} }}

\begin{abstract}
We perform the first global QCD analysis of parton distribution functions (PDFs) in the pion, with lattice-QCD data on gluonic pseudo--Ioffe-time distributions fitted simultaneously with experimental Drell-Yan and leading neutron electroproduction data.
Inclusion of the lattice results with parametrized systematic corrections significantly reduces the uncertainties on the gluon PDF at parton momentum fractions $x \gtrsim 0.2$, revealing a higher gluon density in the pion at large $x$ than in the proton.
The similar gluon momentum fractions in the pion and proton further suggests a relative suppression of the pion gluon density at small $x$.
\end{abstract}

%%%%%%%%%%%%%%%%%%%%%%%%%%%%%%%%%%%%%%%%
\preprint{MSUHEP-25-014, JLAB-THY-25-4417}
\date{\today}
\maketitle
%%%%%%%%%%%%%%%%%%%%%%%%%%%%%%%%%%%%%%%%%%%%%%%%%%%%%%%%%%%%%%%%%%%%%%%%%%%%%%%%

%%%%%%%%%%%%%%%%%%%%%%%%%%%%%%%%%%%%%%%%%%%%%%%%%%%%%%%%%%%%%%%%%%%%%%%%%%%%%%%%
{\it Introduction.---}
As the lightest hadronic bound state in quantum chromodynamics (QCD), the pion displays unique physical characteristics at both low and high energy scales.
On the one hand, it presents itself as the nearly massless pseudo--Nambu-Goldstone boson associated with dynamical chiral symmetry breaking, which is fundamental to understanding hadronic interactions and mass generation at low energies.
On the other hand, its internal quark and gluon structure can be resolved by high energy probes, revealing intriguing differences from that of other hadrons, such as the proton.

Because of the pion's short lifetime ($\tau_{\pi^\pm} \approx 26$~ns), elucidating its substructure, particularly its gluonic component, presents a singular challenge for experiment.
In practice, Drell-Yan (DY) lepton-pair production data from secondary pion beams scattering on nuclear targets~\cite{NA10:1985ibr, Conway:1989fs} has served as the main inputs into global QCD analyses of the pion's parton distribution functions (PDFs)~\cite{Owens:1984zj, Aurenche:1989sx, Sutton:1991ay, Gluck:1991ey, Gluck:1999xe, Wijesooriya:2005ir, Aicher:2010cb, Barry:2018ort, Barry:2021osv, Cao:2021aci, Kotz:2023pbu, Kotz:2025lio}, constraining primarily the pion's valence quark structure at moderate and large parton momentum fractions, $x$.
More recently, leading neutron (LN) electroproduction in deep-inelastic scattering (DIS)~\cite{H1:2010hym, ZEUS:2002gig}, covering a large range of four-momentum transfer squared $Q^2$ has been used to provide constraints on the sea and gluon PDFs at small~$x$~\cite{Barry:2018ort, Barry:2021osv, Cao:2021aci}.

Reduction of uncertainties on the gluon PDF at large~$x$ can be obtained from the transverse momentum dependence of DY lepton pairs, although the impact of the existing data is relatively modest~\cite{Cao:2021aci}, and at present the behavior of the gluon PDF remains essentially unknown at $x \gtrsim 0.1$.
In fact, some recent analyses~\cite{Kotz:2023pbu, Kotz:2025lio} of the DY data and a restricted set of LN data have suggested that more general PDF parametrizations may be consistent with a zero gluon distribution at the input scale.

The current gap in our knowledge in the gluon sector may be partially filled by theoretical advances made over the past decade in lattice-QCD calculations of PDFs mostly using large momentum effective theory~\cite{Ji:2013dva, Ji:2014gla}, good lattice cross sections~\cite{Ma:2014jla, Ma:2017pxb}, or the pseudo-PDF~\cite{Radyushkin:2017cyf} frameworks to connect Euclidean correlators with PDFs through perturbative matching.
The pseudo-PDF method, which relies on a short distance factorization, has proven to be effective for calculating gluon PDFs, whose correlators have limited signal at large distances~\cite{Fan:2018dxu, Fan:2020cpa, Fan:2021bcr, HadStruc:2021wmh, Salas-Chavira:2021wui, Fan:2022kcb, Delmar:2023agv, Good:2023ecp, HadStruc:2022yaw, Khan:2022vot, Karpie:2023nyg, NieMiera:2025inn}.
Recent progress has been made in incorporating  lattice-QCD valence-quark pion  data into global QCD analyses to provide additional constraints on PDFs in kinematic regions that are difficult to access experimentally, quantifying both the uncertainties of the pion PDFs and systematic effects intrinsic to the lattice-QCD observables~\cite{JeffersonLabAngularMomentumJAM:2022aix}.

In this Letter, we report on a new global QCD analysis of PDFs in the pion, in which the experimental DY and LN observables are supplemented by recent lattice calculations of reduced pseudo--Ioffe-time distributions (RpITDs)~data~\cite{Good:2024iur}.
The resulting simultaneous analysis, including a rigorous treatment of lattice systematic effects, significantly decreases the uncertainties on the gluon PDF in the pion, allowing quantitative comparisons to be made with the more precisely determined gluon PDFs in the proton.

%%%%%%%%%%%%%%%%%%%%%%%%%%%%%%%%%%%%%%%%%%%%%%%%%%%%%%%%%%%%%%%%%%%%%%%%%%%%%%%%
{\it Lattice data.---} 
The RpITD is defined by a double ratio of the matrix elements between pion states, with momentum $P_z$, of gluonic correlators with spatial separation in the $z$-direction, $M(z, P_z) = \langle \pi(P_z) | O(z) | \pi(P_z) \rangle$, which can be related via a short distance factorization to the gluon PDF $g(x)$ and quark-singlet PDF $\Sigma(x)$~\cite{Balitsky:2019krf, Balitsky:2021qsr, MorrisIII:2022fav},
\begin{eqnarray}
     && \hspace*{-0.5cm}
\mathcal{M}(\nu, z^2)  
\equiv \frac{M(z, P_z )/M(z, 0)}{M(0, P_z)/M(0, 0)} 
     \notag\\ 
     && \hspace*{-0.5cm}
= \int_0^1 \dd x \frac{xg(x)}{\langle x_g \rangle}
  \left[R_{gg}(x\nu, z^2) 
      + R_r(x\nu, z^2) \frac{\langle x_\Sigma \rangle}{\langle x_g \rangle} \right]  
      \notag\\
     && \hspace*{-0.5cm}
+ \int_0^1 \dd x \frac{x\Sigma(x)}{\langle x_g \rangle} R_{gq}(x\nu, z^2) 
%+ \mathcal{O}(m^2z^2, z^2 \Lambda_\text{QCD})
+ z^2 B_1(\nu) + \frac{1}{z}P_1(\nu),
\label{eq:matching}
\end{eqnarray}
where $\langle x_g \rangle$ and $\langle x_\Sigma \rangle$ are the corresponding  momentum fractions, and $\nu = z P_z$ is the Ioffe time.
The ratio removes the renormalization and kinematic factors and suppresses lattice systematic effects~\cite{Radyushkin:2017cyf, Orginos:2017kos}.
In Eq.~(\ref{eq:matching}) the perturbatively calculated kernel $R_{gg}$ is the gluon-gluon matching term, $R_{gq}$ is the gluon-quark mixing kernel, and $R_r$, which was neglected in earlier derivations of the matching~\cite{Balitsky:2019krf}, ensures normalization of the RpITD in the presence of gluon-quark mixing.
At leading order, Eq.~(\ref{eq:matching}) reduces to a Fourier transform of $xg(x)/\langle x_g \rangle$, revealing direct sensitivity to the shape of the gluon PDF.
The next-to-leading order (NLO) corrections provide a modest correction to the matching, with the quark mixing contribution being negligible compared to statistical uncertainties in the lattice data~\cite{Fan:2021bcr, Fan:2022kcb, Delmar:2023agv}, in part due to the color factors %$C_A=3$ versus $C_F=4/3$ 
on the gluon NLO and quark mixing terms.

In addition to the leading-twist contributions in the matching, the terms $z^2 B_1(\nu)$ and $(1/z)P_1(\nu)$ account for systematic higher-twist and discretization effects~\cite{Karpie:2021pap, JeffersonLabAngularMomentumJAM:2022aix}, respectively, and can be parameterized by
\begin{align}\label{eq:sys_param}
    B_1(\nu) &= \sum_{n=1}^N b_n\, \sigma_{0,n}(\nu), \quad
    P_1(\nu)  = \sum_{n=1}^N p_n\, \sigma_{0,n}(\nu), 
\end{align}
where $\sigma_{0,n}$ are the weighted cosine transforms of Jacobi polynomials, shifted to the domain $0\leq x \leq 1$,
\begin{equation}\label{eq:sigma_def}
    \sigma_{0,n}(\nu) = \int_0^1 \dd{x}\, \cos(\nu x)\, x^a (1-x)^b\, J_n^{(a,b)}(x).
\end{equation}
This construction ensures that $\sigma_{0,n}(0) = 0$, which is required for proper normalization of the RpITD.
The parameters $\{ a, b, b_n, p_n \}$ are optimized during the global fit, and for any $a, b > -1$, as $N \to \infty$, $B_1$ and $P_1$ can reproduce any (even) function of Ioffe time.

In practice we truncate the sum in Eqs.~(\ref{eq:sys_param}) to $N=2$ \cite{Karpie:2021pap, JeffersonLabAngularMomentumJAM:2022aix}, and use a normal distribution centered at 0 with a width of 0.75 as a Bayesian prior on $a$ to improve convergence of Eq.~\ref{eq:sigma_def}.

The RpITD data~\cite{Good:2024iur} were obtained from
approximately $1.3$\,M two-point correlators measured across 1,013 configurations with lattice spacing $a = 0.1207$\,fm at a valence pion mass $M_\pi \approx 310$\,MeV, generated by the MILC collaboration~\cite{MILC:2013znn} using 2+1+1 flavors of highly-improved staggered quarks (HISQ)~\cite{Follana:2007rc} with a lattice volume of $24^3 \times 64$.
Wilson-clover fermions were used in the valence sector, with quark masses tuned to reproduce the light and strange masses of the HISQ sea.
Gaussian momentum smearing~\cite{Bali:2016lva} was used on the quark field to improve the signal for boost momenta $P_z$ up to 2.2~GeV.

For the three-point correlators, five steps of hypercubic smearing~\cite{Hasenfratz:2001hp} were used to obtain reliable signals up to $z = 9a$.
It is not possible to fit the $z = a$ and $2a$ results due to discretization effects that cannot be accurately accounted for through the systematic term $1/\big(z P_1(\nu)\big)$, which leaves 35 lattice RpITD data inputs from 5 values of $P_z \in [0.43, 2.2]$~GeV and 7 values of $z \in [3a,9a]$.
We use the matching scale $\mu^2 = \text{max}\left[m_c^2, 4/(z^2 e^{2\gamma_E})\right]$, to optimize the perturbative convergence and maintain applicability of perturbation theory.

%%%%%%%%%%%%%%%%%%%%%%%%%%%%%%%%%%%%%%%%%%%%%%%%%%%%%%%%%%%%%%%%%%%%%%%%%%%%%%%%
{\it Global analysis.---} 
For the experimental data we follow the previous JAM analysis~\cite{Barry:2021osv},  which fitted pion-nucleus DY cross sections from the NA10~\cite{NA10:1985ibr} (56 data points) and E615~\cite{Conway:1989fs} (61 points) collaborations, employing 
NLO plus next-to-leading log (NLL) perturbative corrections with the double Mellin method.
For the LN electroproduction data at small $x$ from HERA, we use the LN cross sections from H1~\cite{H1:2010hym} (58 points) and the more precise ratio of LN to inclusive cross sections from ZEUS~\cite{ZEUS:2002gig} (50 points), which are expected to be dominated by single pion exchange for small neutron scattering angles and large longitudinal momentum fractions~\cite{Sullivan:1971kd, Melnitchouk:1995en, McKenney:2015xis}.

Following the standard paradigm adopted by the JAM collaboration, we parameterize the valence quark ($q_v$), sea quark ($q_s$), and gluon ($g$) distributions at an initial scale $\mu_0$, using the common template function
\begin{equation}\label{eq:parmeterization}
    f(x,\mu_0) = \frac{N\, x^{\alpha}(1-x)^{\beta}(1+\gamma x^2)}{B(\alpha+2, \beta+1) + \gamma B(\alpha+4,\beta+1)}
\end{equation}
for each parton flavor $f$, setting the initial scale to the charm-quark mass, $\mu_0 = m_c = 1.28$~GeV, in the $\overline{\text{MS}}$ scheme.
Charge symmetry imposes constraints on the valence-quark PDFs,
    $q_v \equiv \bar{u}_v^{\pi^-} \equiv \bar{u}^{\pi^-}-u^{\pi^-} = d_v^{\pi^-}$,
and we adopt a flavor-symmetric quark sea,
    $q_s \equiv u^{\pi^-} = \bar{d}^{\pi^-} = s^\pi = \bar{s}^\pi$.
The normalizations of the valence and sea quarks are constrained at the model scale by baryon-number conservation,
    $\int_0^1 \dd{x} q_v = 1$,
and the momentum sum rule,
    $\int_0^1 \dd{x} x\, ( 2 q_v + 6 q_s + g) = 1$,
which are satisfied at all scales $\mu$ by DGLAP evolution.
The weak sensitivity to the $\gamma$ parameter for the sea quark and gluon PDFs allows us to set $\gamma=0$ for these, and choosing the normalizations $N$ to be positive and $\gamma > -1$ for the valence PDF restricts all PDFs to be positive.

We use Bayesian statistics to sample the posterior distribution of the PDF and lattice systematic parameters using data resampling approach according to 
    $\mathcal{P}(\mathbf{a}|\text{data}) 
    \propto \mathcal{L}(\text{data}|\mathbf{a})\, \pi(\mathbf{a})$,
where $\pi(\mathbf{a})$ is the prior distribution of the parameters, and 
    $\mathcal{L}(\text{data}|\mathbf{a}) 
    = \exp\left[-\frac{1}{2}\chi^2(\mathbf{a},\text{data})\right]$
is the likelihood.
The sum of the experimental and lattice $\chi^2$ values, $\chi^2 = \chi^2_\text{exp} + \chi^2_\text{lat}$, is minimized using the procedure in Ref.~\cite{JeffersonLabAngularMomentumJAM:2022aix}.
Since the lattice data are strongly correlated, $\chi^2_\text{lat}$ incorporates the statistical covariance matrix.
The posterior distributions are sampled by performing $N_\text{rep}$ $\chi^2$ minimizations to bootstrapped data to obtain $N_\text{rep}$ replicas.
For each replica, the experimental data are reshuffled with a Gaussian distribution for the uncorrelated uncertainties, and the lattice data according to a multivariate normal distribution within the covariance matrix.

Our baseline fit is to the experimental DY and LN data using fixed-order NLO with threshold resummation for DY~\cite{Barry:2021osv}.
The new fits incorporating the lattice data include leading twist (LT) only RpITD matching for the lattice data, and the full matching with the systematic terms $B_1$ and $P_1$ in Eq.~(\ref{eq:matching}).
The quality of fit is assessed by computing the reduced $\overline{\chi}^2 \equiv \chi^2/N_\text{dat}$ for the experimental and lattice data, and the $Z$-scores, $Z = \sqrt{2}\, \text{erf}^{-1}(1-2p)$, where $p$ is the $p$-value estimated from a $\chi^2$ distribution with $N_\text{dat}$ degrees of freedom.

%%%%%%%%%%%%%%%%%%%%%%%%%%%%%%%%%%%%%%%%%%%%%%%%%%%%%%%%%%%%%%%%%%%%%%%%%%%%%%%%
{\it Fit results.---}
We perform our global analysis over ${\cal O}(800)$ replicas to achieve convergence of statistical quantities.
We find for the experimental data a reduced $\overline{\chi}^2_\text{exp} = 0.86$ before and after the inclusion of lattice data, with or without the systematic terms in the lattice matching, suggesting that the lattice data have no impact on the fit to the experimental data.
Including the lattice data without the systematic terms gives relatively large reduced $\overline{\chi}^2_\text{lat}$ ($Z$-score) values of 1.86 (2.96).
With the systematic terms included, the $\overline{\chi}^2_\text{lat}$ ($Z$-score) decreases to a much more acceptable 1.25 (1.05).
This is in agreement with previous findings \cite{JeffersonLabAngularMomentumJAM:2022aix} that systematic terms are important when simultaneously fitting experimental and lattice data for pion valence-quark PDFs.
In the subsequent analysis we will treat the fit {\it with} systematics as our main result with lattice data.

\begin{figure}[tbp]
\hspace*{-0.2cm}
\includegraphics[width=.47\textwidth]{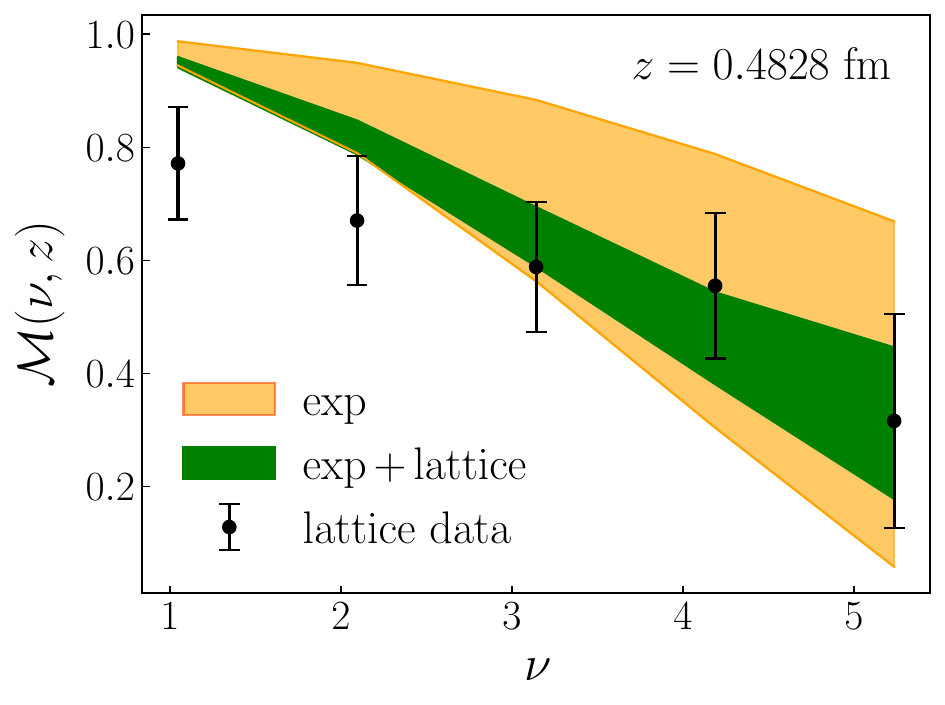}
\hspace*{-0.2cm}
\includegraphics[width=.47\textwidth]{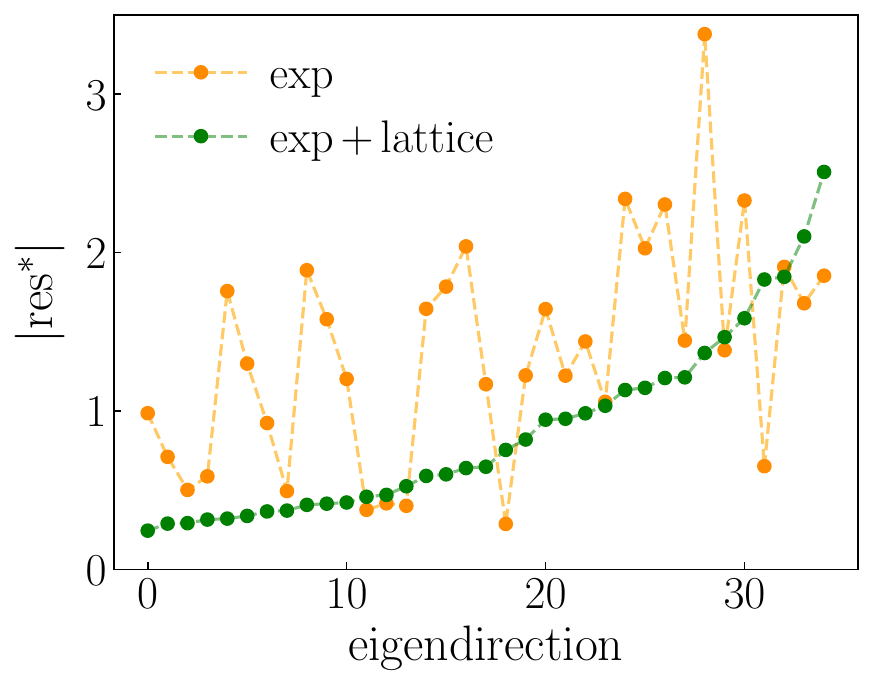}
    \caption{({\bf Upper}) Lattice RpITD data at $z = 0.4828$~fm (black circles) compared with reconstructed ratios from the fits with (green band) and without (orange band) lattice data.
    ({\bf Lower}) Absolute value of the residuals in the eigendirections of the covariance matrix with (green circles) and without (orange circles) the lattice data.
    }
    \label{fig:RpITD}
\end{figure}

%\BG{RpITD plots}
To illustrate the comparison between the fitted results and the lattice RpITD data, in Fig.~\ref{fig:RpITD} we plot the RpITD ${\cal M}(\nu,z)$ for a representative value $z = 4a = 0.4828$~fm as a function of Ioffe time.
The reconstructed shape from the experiment-only fit is consistent with the general trend of the RpITD data within uncertainties, which are significantly reduced when the lattice data are combined with experiment.
The agreement of the reconstructed ratio is generally better at larger $\nu$ (or $P_z$), which may suggest systematic contamination in the lattice data which scales with $1/P_z$.

On the other hand, it should be noted that the lattice data are highly correlated, and a simple data versus theory comparison may not accurately convey the full uncertainties.
This can be addressed by diagonalizing the covariance matrix, ${\bm \Sigma} = \mathbf{U}^T\mathbf{D}\mathbf{U}$, to write the lattice $\chi^2$ as a sum of independent residuals in the eigendirections of the covariance matrix~\cite{Karpie:2023nyg},
\begin{align}
  \chi_\text{lat}^2 &= (\mathbf{d}-\mathbf{t})^T \mathbf{\Sigma}^{-1}(\mathbf{d}-\mathbf{t})
   = (\textbf{res}^*)^2,
\end{align}
where $\mathbf{d}$ is the vector of lattice RpITD data at every kinematic point, $\mathbf{t}$ is the RpITD calculated from our analysis, and $\textbf{res}^* = \mathbf{D}^{-1/2}\mathbf{U}(\mathbf{d}-\mathbf{t})$.
In the lower panel of Fig.~\ref{fig:RpITD} we plot the mean $\textbf{res}^*$ for each of the 35 eigendirections for each fit ordered from smallest to largest with respect to the combined fit.
The fit to the lattice matrix elements reduces the majority of the residuals relative to the fit without lattice.
However, the lattice kinematics are spread across the eigendirections such that there is no clear way to connect individual residuals to specific kinematic regions.

\begin{figure}[t]
\hspace*{-0.3cm}
\includegraphics[width=0.5\textwidth]{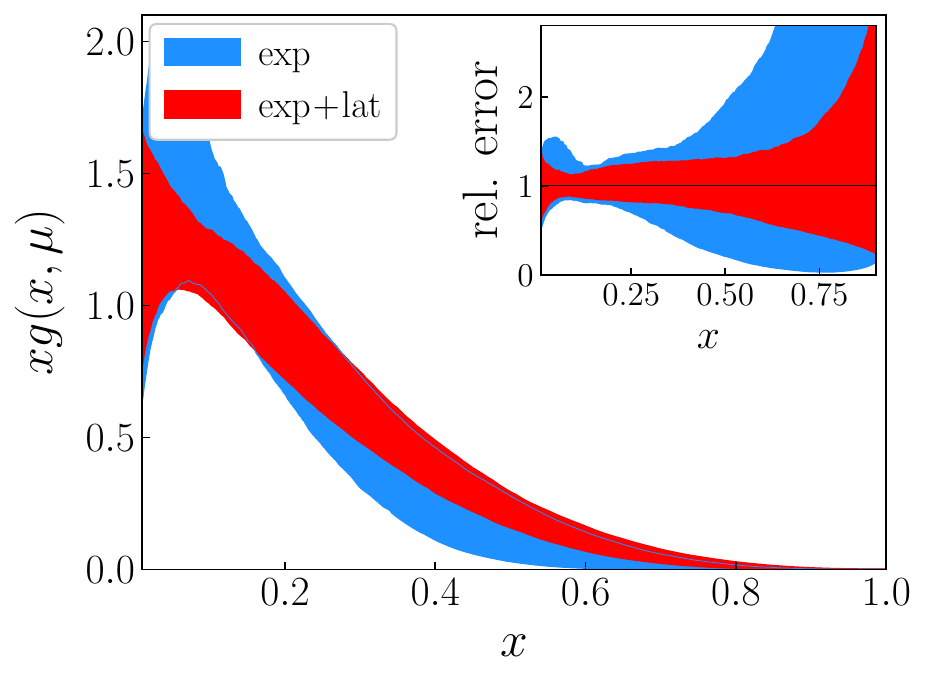}
    \caption{Gluon PDF in the pion, $x g(x,\mu)$, at the input scale $\mu = m_c$ with 68\% credibility interval bands for the fits before (blue) and after (red) inclusion of lattice data. The inset shows the uncertainty bands relative to the median PDFs.}
    \label{fig:glue_PDFs}
\end{figure}

\begin{figure}[t]
\hspace*{-0.3cm}
\includegraphics[width=0.5\textwidth]{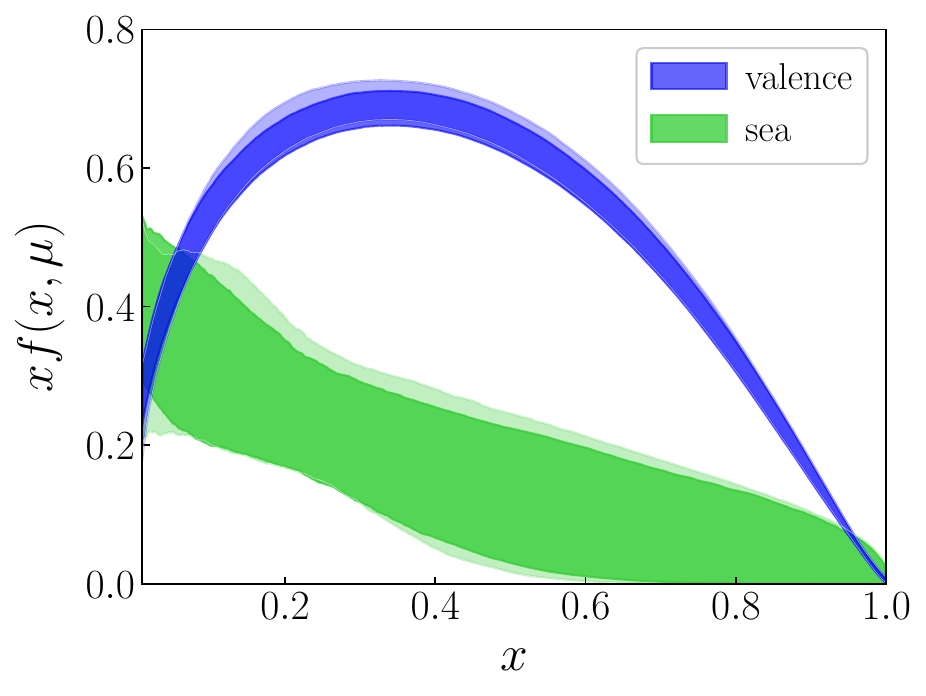}
    \caption{68\% credibility interval bands for the valence (blue) and sea (green) quark PDFs at $\mu = m_c$ with 68\% credibility interval for the fits before (lighter bands) and after (darker bands) inclusion of lattice data.}
\label{fig:quark_PDFs}
\end{figure}

The effect of including lattice data on the gluon PDF is illustrated in Fig.~\ref{fig:glue_PDFs} at the input scale $\mu=m_c$.
The lattice data drive the gluon PDF to be slightly larger at high $x$, while at small $x$ it retains a more similar size and shape.
Furthermore, the PDF with lattice data is still within the uncertainties of the fit without lattice data.
A significant reduction in the gluon PDF uncertainties is observed, exceeding $50\%$ at $x \gtrsim 0.5$.
On the other hand, the impact on the valence- and sea-quark PDFs, illustrated in Fig.~\ref{fig:quark_PDFs}, is very small and lie within the uncertainties;
this is expected from the weak dependence of the calculated RpITD on the quark PDFs.

\begin{figure}[t]
\hspace*{-0.2cm}\includegraphics[width=0.5\textwidth]{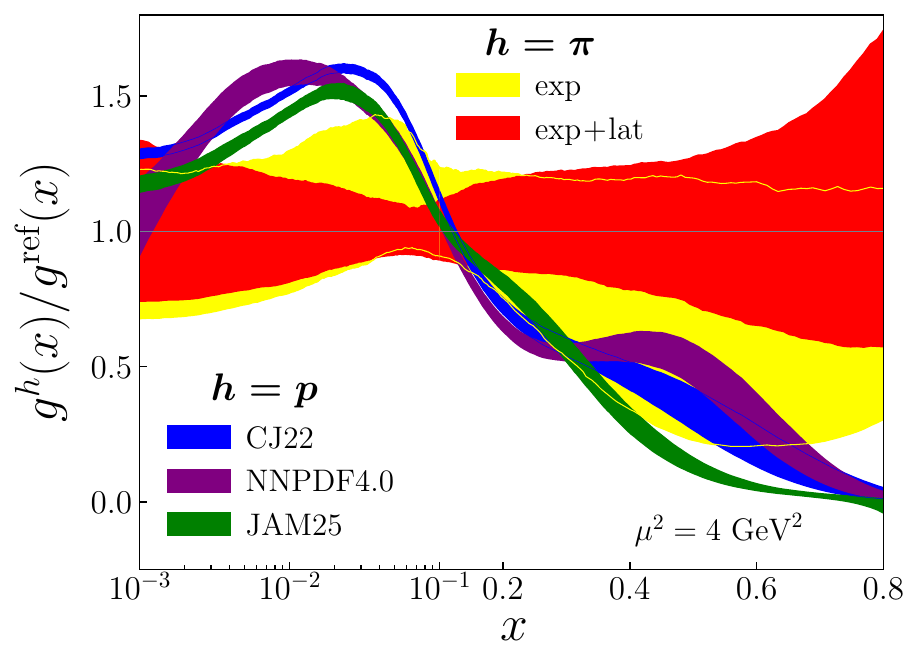}
\caption{Gluon PDFs $g^h$ in pions ($h=\pi$) and protons ($h=p$), relative to the reference gluon pion PDF $g^\text{ref}$ given by the central value of the pion PDF fit, for the analysis before (yellow band) and after (red) inclusion of lattice data. The gluon PDFs in the proton from the CJ22~\cite{Cerutti:2025yji} (blue), NNPDF4.0~\cite{NNPDF:2021njg} (purple), and JAM25~\cite{Cocuzza:2025qvf} (green) parametrizations are shown for comparison. All bands represent 68\% credibility intervals.}
\label{fig:proton-pion}
\end{figure}

The improved determination of the gluon PDF in the pion allows for the first time a quantitative comparison with the gluon PDF in the proton.
In Fig.~\ref{fig:proton-pion} we show a sample of gluon PDFs for the proton, from the CJ22~\cite{Cerutti:2025yji}, NNPDF4.0~\cite{NNPDF:2021njg}, and JAM25~\cite{Cocuzza:2025qvf} global analyses, and the pion divided by the central value of the pion gluon PDF in the full (exp+lat) fit at $\mu^2 = 4$~GeV$^2$.
% We can see immediately that the gluon distributions in the hadrons have quite different shapes.
Prior to the addition of the lattice constraints, the experimental data provided sensitivity to the $g^\pi$ PDF for $x \lesssim 0.2$, so that differences between gluons in the pion and proton could only be identified at $x \approx 0.003 - 0.04$.
With the addition of lattice data and consequent reduction of the pion PDF uncertainties, the gluon pion and proton PDFs become more distinct in both the small-$x$ ($x \approx 0.01-0.1$) and large-$x$ ($x \gtrsim 0.2$) regions.
In particular, one observes a clear enhancement of $g^\pi$ compared with $g^p$ at large $x$, which is consistent with general expectations from power counting behaviors, although the difference appears greater than the predicted single power $(1-x)$~\cite{Courtoy:2020fex}.

Interestingly, the respective low-$x$ and high-$x$ suppression and enhancement balance each other, such that the  gluon momentum fractions in the pion and proton are very similar.
At $\mu = 2$~GeV, we find the pion gluon momentum fraction $\langle x \rangle_g^\pi = 0.43(4)$, while the fractions for the proton are in the range  $\langle x \rangle_g^p \approx 0.40$--0.42~\cite{Cerutti:2025yji, NNPDF:2021njg, Cocuzza:2025qvf, Bailey:2020ooq, Hou:2019efy}.
This has important implications for the decomposition of the pion and proton masses.
In particular, the fraction of the gluonic mass in a hadron is $\tfrac34 \langle x \rangle_g$~\cite{Ji:1994av}, so that despite the large disparity between the pion and proton masses, the fractions carried by gluons are both about $30$\%, which may point to a common origin of both.

% \vspace*{0.3cm}

%%%%%%%%%%%%%%%%%%%%%%%%%%%%%%%%%%%%%%%%%%%%%%%%%%%%%%%%%%%%%%%%%%%%%%%%%%%%%%%%
{\it Outlook.---}  
Our central finding of a significant reduction in uncertainties on the gluon PDF in the pion with the inclusion of the lattice RpITD data is a clear indication of a higher density of gluons in the pion than in the proton at large $x$, and consequently a suppression in the pion gluon density at small $x$.
Future lattice data with higher statistics and other signal improvement methods will further improve the constraining power of the gluon RpITDs, and inclusion of quark flavor-singlet data will provide additional constraints on the sea quark PDFs.
On the experimental front, the planned COMPASS++ and AMBER facilities at CERN~\cite{Adams:2018pwt} along with the leading baryon production measurements in tagged DIS at Jefferson Lab~\cite{TDIS} should provide direct constraints of quark pion PDFs at intermediate and high $x$. 
Similar tagged DIS measurements at future electron-ion colliders~\cite{Achenbach:2023pba, Anderle:2021wcy} will further probe the structure of the pion sea-quark and gluon distributions at small $x$. \\

% \clearpage
%%%%%%%%%%%%%%%%%%%%%%%%%%%%%%%%%%%%%%%%%%%%%%%%%%%%%%%%%%%%%%%%%%%%%%%%%%%
%\begin{acknowledgments} 
{\it Acknowledgments.---}\  
We thank the MILC Collaboration for sharing the lattices used to perform this study.
The LQCD calculations were performed using the Chroma software suite~\cite{Edwards:2004sx}. 
This research used resources of the National Energy Research Scientific Computing Center, a DOE Office of Science User Facility supported by the Office of Science of the U.S. Department of Energy under Contract No. DE-AC02-05CH11231 through ERCAP;
facilities of the USQCD Collaboration, which are funded by the Office of Science of the U.S. Department of Energy,
and supported in part by Michigan State University through computational resources provided by the Institute for Cyber-Enabled Research (iCER).  
The work of WG and AN is partially supported by the U.S. Department of Energy, Office of Science, under grant DE-SC0024053 ``High Energy Physics Computing Traineeship for Lattice Gauge Theory''. 
The work of WG and HL is  partially supported by the US National Science Foundation under grant PHY~2209424. 
The work of PB was supported by the U.S.~Department of Energy, Office of Science, Office of Nuclear Physics, contract no.~DE-AC02-06CH11357, and the Scientific Discovery through Advanced Computing (SciDAC) award {\it Femtoscale Imaging of Nuclei using Exascale Platforms}.
The work of WM and NS was supported by the DOE contract No.~DE-AC05-06OR23177, under which Jefferson Science Associates, LLC operates Jefferson Lab. 
NS was supported by the DOE, Office of Science, Office of Nuclear Physics in the Early Career Program.
%\end{acknowledgments} 

%%%%%%%%%%%%%%%%%%%%%%%%%%%%%%%%%%%%%%%%%%%%%%%%%%%%%%%%%%%%%%%%%%%%%%%%%%%
\bibliography{refs}

\end{document}